\documentclass[10pt]{IETBook}
\usepackage{array}
\usepackage{amsmath}
\usepackage{booktabs}
\usepackage{amssymb}
\usepackage{pifont}
\usepackage{rotating}
\usepackage{lipsum}
\usepackage{setspace}
\usepackage[numbers]{natbib}
\usepackage{authblk}
\usepackage{url}

\usepackage{breakurl}
\usepackage[breaklinks]{hyperref}

\newcolumntype{L}[1]{>{\raggedright\let\newline\\\arraybackslash\hspace{0pt}}m{#1}}
\newcolumntype{C}[1]{>{\centering\let\newline\\\arraybackslash\hspace{0pt}}m{#1}}
\newcolumntype{R}[1]{>{\raggedleft\let\newline\\\arraybackslash\hspace{0pt}}m{#1}}

\begin{document}

%For RH Book title
%\rhbooktitle{Big Data-enabled IoT: Challenges and Opportunities}

%\markboth{Big Data-enabled IoT: Challenges and Opportunities}{Fog Computing Architecture:  Survey and Challenges}

\cauthor{Ranesh Kumar Naha\thanks{School of Technology, Environments and Design, University of Tasmania, Hobart, TAS 7001, Australia Email:raneshkumar.naha@utas.edu.au}
Saurabh Garg \thanks{School of Technology, Environments and Design, University of Tasmania, Hobart, TAS 7001, Australia Email:saurabh.garg@utas.edu.au} and\\
Andrew Chan\thanks{School of Engineering, University of Tasmania, Hobart, TAS 7001, Australia Email:andrew.chan@utas.edu.au}}

\chapter{Fog Computing Architecture:  Survey and Challenges}

\textbf{Abstract:} Emerging technologies that generate a huge amount of data such as the Internet of Things (IoT) services need latency aware computing platforms to support time-critical applications. Due to the on-demand services and scalability features of cloud computing, Big Data application processing is done in the cloud infrastructure. Managing Big Data applications exclusively in the cloud is not an efficient solution for latency-sensitive applications related to smart transportation systems, healthcare solutions, emergency response systems and content delivery applications. Thus, the Fog computing paradigm that allows applications to perform computing operations in-between the cloud and the end devices has emerged. In Fog architecture, IoT devices and sensors are connected to the Fog devices which are located in close proximity to the users and it is also responsible for intermediate computation and storage. Most computations will be done on the edge by eliminating full dependencies on the cloud resources. In this chapter, we investigate and survey Fog computing architectures which have been proposed over the past few years. Moreover, we study the requirements of IoT applications and platforms, and the limitations faced by cloud systems when executing IoT applications. Finally, we review current research works that particularly focus on Big Data application execution on Fog and address several open challenges as well as future research directions. 

\section{Introduction}
IoT is a connected network of things where all connected nodes are constantly communicating together automatically in coordination to produce collective results in order to serve people for a better life and economic advancement. It promises to serve everyone to get access to any service or network related to objects, people, programs and data from anywhere. IoT also enables communication between machines, people and things to implement an environment for smart living, smart cities, smart transportation systems, smart energy distribution, smart health services, smart industries, smart buildings, and they all work together towards making this planet a smarter planet \cite{vermesan2014internet,tammishetty2017iot}. 

From the IoT environment, Big Data are generated in each and every moment from sensors, messaging systems, mobile devices and social networks resulting in a new form of network architecture. There are many technical challenges related to the IoT environment arising due to its distributed, complex and dynamic nature. These technical challenges include connectivity, capacity, cost, power, scalability and reliability \cite{vermesan2014internet}. Another key issue is the processing of Big Data that is produced from various IoT nodes. Generally, we rely on the cloud to process Big Data but sometimes it is really not feasible to transfer all generated data to the cloud for processing and storage. The process of sending all generated data to the cloud might occupy a certain amount of network bandwidth and on the other hand, the cloud is not able to process latency aware applications due to high response times. Hence, Fog computing came into the picture to process Big Data near to user locations.

\begin{figure}[!b]
	\centering
	\includegraphics[width=3.5in]{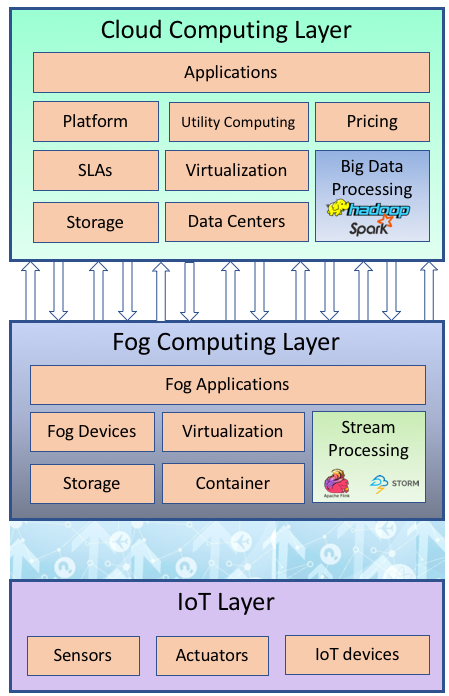}
	%where an .eps filename suffix will be assumed under latex, 
	%and a .pdf suffix will be assumed for pdflatex; or what has been declared
	%via \DeclareGraphicsExtensions.
	\caption{Flow and processing of Big Data in IoT environment.}
	\label{bdflowfog}
\end{figure}

Processing Big Data from the Fog environment is an emerging computing paradigm which brings application services closer to the users with better service quality. Any device that has converged infrastructure can act as a Fog node and provide computation, storage, and networking services to the users. Fog devices are also connected to the cloud to handle complex processing and long-term storage, and this computing paradigm also referred as IoT-Fog-cloud framework \cite{yousefpour2018reducing}. Figure \ref{bdflowfog} shows the flow and the processing of Big Data in the IoT environment.

\section{Fog computing architecture}
Fog computing is designed to be deployed in a distributed manner where edge devices do the processing. In contrast, cloud computing is a more centralized concept. In Fog, processing and storage devices are located in close proximity compared to the cloud and this is the reason why Fog is more capable to serve latency aware services through access points, smart phones, base stations, switches, servers, and routers. The services which are referred to as low latency services are mostly emergency services including natural disasters, healthcare and so on. Apart from that, augmented reality, video streaming, gaming and any other smart communication system also requires time- sensitive computation. With regards to improving quality of life via technology, Fog computing is going to play a major role in the near future. In this section, we discuss several current research studies on Fog computing architecture before presenting a high-level Fog computing architecture.

\subsection{Existing research on Fog computing architecture}
As Fog computing has emerged recently, no standard architecture is available so far for Fog computing paradigms \cite{aazam2015fog}. Several studies \cite{aazam2015fog, arkian2017mist, luan2015fog, giang2015developing, dastjerdi2016fog, openfog2016openfog, intharawijitr2016analysis, nadeem2016fog, taneja2016resource, hosseinpour2016approach, baccarelli2017fog, sun2017resource, munir2017ifciot}  have proposed various architectures of Fog computing. The first Fog computing architecture was depicted by Bonomi et al. \cite{bonomi2012fog} where the Fog layer was defined as distributed intelligence which resides in between the core network and sensor devices. Bonomi et al. \cite{bonomi2012fog} also point-out several characteristics which makes Fog a non-trivial extension of cloud computing. These characteristics are edge location, low latency, massive sensor network, very large number of nodes, mobility support, real-time interaction, dominant wireless connectivity, heterogeneity, interoperability, distributed deployment, on-line analytics and interplay with the cloud. Several types of architecture such as layer-based, hierarchical and network-based are proposed by various researchers for Fog computing as described below.

\subsubsection{Fog layered architecture} Aazam et al. \cite{aazam2015fog} presented a layered architecture of Fog where six different layers were shown. Physical and virtual nodes as well as sensors were maintained and managed according to their service types and requirements by the lower layer known as the physical and virtualization layer. The next upper layer is the Monitoring layer which monitors network and underlying node activities. This layer defines when and what task needs to be performed by which node and also takes care of the subsequent task requirements that need be executed to complete the main task. The same layer also monitors power consumption for energy constrained devices or nodes. Above the monitoring layer, the pre-processing layer resides which performs data management related tasks to get necessary and more meaningful data. After that, data is stored temporarily in the Fog resources by the next upper layer known as the temporary storage layer. Once the processed data is uploaded to the cloud, these are removed from the local data storage media. For private data, the security layer provides privacy, encryption and integrity measure related service. The topmost layer uploads pre-processed and secured data to the cloud. In this way, most processing will be done in the Fog environment and allows the cloud to deal with more complex services. 

Arkian et al. \cite{arkian2017mist} proposed a four layer Fog architecture: i) Data generator layer, ii) Cloud computing layer, iii) Fog computing layer and iv) Data consumer layer. A wide range of consumers is considered from individuals to enterprises in the data consumer layer. Consumers can submit their requests to three other layers and get responses for the required services. The data generator layer is where the IoT devices reside and communicate with the cloud computing layer via the Fog computing layer. In this architecture, data pre-processing will be done in the Fog computing layer. This layer also enables context awareness and low latency. The cloud computing layer provides centralized control and a wide range of monitoring services. Long-term and complex behaviour analysis will be performed at this layer to support dynamic decision making such as relationship modelling, long-term pattern recognition, and large-scale event detection. The key difference between this architecture with others above is the direct communication between consumers and all three layers.  

The fog layer is presented as an intermediate layer between mobile devices and the cloud in the Fog system architecture by Luan et al. \cite{luan2015fog}. According to this architecture, the main component of the Fog layer is the Fog server, which should be deployed at a fixed location on the local premises of the mobile users. A Fog server could be an existing network component like a base station or WiFi access point. These servers communicate with mobile devices through single-hop wireless connections and provide them with pre-defined application services in its wireless coverage without seeking assistance from the cloud or other fog servers. This system architecture does not consider many other aspects but discusses the idea of a Fog server.   

Dastjerdi et al. \cite{dastjerdi2016fog} presented a five layer Fog computing reference architecture. The topmost layer is the IoT application layer which provides application facilities to the end users. The next lower layer is the software-defined resource management layer which deals with monitoring, provisioning, security, and management related issues. The next following layer is responsible for managing cloud services and resources. The next lower layer is the network layer which works to maintain the connectivity between all devices and the cloud. The bottommost layer consists of end devices such as sensors, edge devices, and gateways. It includes some apps which belong to this layer and these apps work by improving the device functionality. In this reference architecture, the Fog layer is completely absent and it also did not testify where the computation is done. 

A high-level architecture focusing on the communication in the Fog environment is shown by Nadeem et al. \cite{nadeem2016fog} where communication among devices is shown in three different layers. A similar conceptual architecture was also presented by Taneja et al. \cite{taneja2016resource} and Sarkar et al. \cite{sarkar2016theoretical} where the devices are located in three different tiers. But Taneja et al. \cite{taneja2016resource} and Sarkar et al. \cite{sarkar2016theoretical} added gateway devices to connect the Fog devices and the cloud. These gateway devices are located in tier 2 which is also denoted as the Fog layer.

\subsubsection{Hierarchical Fog architecture} Giang et al. \cite{giang2015developing} presented a hierarchical Fog architecture by classifying Fog devices into three different types based on their processing resources edge, computing, and input/output nodes. Edge nodes execute actuation messages and generate sensing data. Input-Output (IO) nodes have limited computing resources and maintain brokering communications with Edge nodes. Compute nodes have some computing resources to offer and this node is dynamic with a programmable runtime. These three nodes can be implemented separately or with a combination based on system designer preference. 

A conceptual hierarchical architecture of Fog computing is presented by Hosseinpour et al. \cite{hosseinpour2016approach} where the Fog computing layer is divided into three basic levels and can be extended to N numbers of levels. Computation and storage are done at all levels except the lowermost level. Level 0 consists of sensors and actuators, level 1 is named as a gateway Fog node and level 2 represents the core Fog nodes.

\subsubsection{OpenFog architecture} 
The OpenFog architecture explanation is the most comprehensive one in which most Fog computing characteristics were considered \cite{openfog2016openfog}. However, OpenFog architecture did not consider lower latency storage facilities near to business deployment and end users. This architecture intends to do computation near to the end user to minimize latency, migration costs and other network related costs along with bandwidth costs. Without synchronizing and routing all communication to the core network, low latency communication can take place and user requests are routed to the location closest to the end-users where computation elements are available. The implementation of management elements, including configuration and control management, and network measurements are deployed near to the endpoint rather than being controlled from the gateway. In addition, the proposed architecture allows collection and processing of data using local analytics and the results are copied to the cloud in a secure manner for further processing and future use. Although they covered the maximum number of aspects about the Fog computing environment in their Fog computing architecture explanation, there is a lack of proper validation of their described architecture through experimental deployment. However, they intend to collaborate with various related groups, including but not limited to, the Industrial Internet Consortium (IIC), ETSI-MEC (Mobile Edge Computing), Open Connectivity Foundation (OCF) and the Open Network function virtualization (OpenNFV).

\subsubsection{Fog network architecture} 
Intharawijitr et al. \cite{intharawijitr2016analysis} illustrated the Fog network architecture and considered communication latency issues faced by Fog devices connected via a 5G cellular network. In their model, the edge router works as a Fog server and does the processing for the users as mobile users send packets to the Fog server. The Fog server does not pass the request to the core network unless the request is for a cloud service. A mathematical model was defined to clarify communication delays and computing delays in the Fog network and other related parameters. Three different policies were defined to choose a target Fog server for every task. To validate the proposed model, an experimental evaluation was carried out in a simulation environment by employing the proposed policies.

\subsubsection{Fog architecture for Internet of energy} An envisioned Fog of Everything (FoE) technology platform was proposed by Baccarelli et al. \cite{baccarelli2017fog}. In the architecture of this platform, all Fog devices (IoT sensors, smart car, smartphone or any other station) will be connected to the wireless base station via Fog to Things (F2T) and Things to Fog (T2F) two-way connectivity through TCP/IP connections functioning onto IEEE802.11/15 single-hop links. All fog devices connected with the same base station are considered to be in the same cluster. All base stations are considered as Fog nodes and will be connected via Fog to Fog (F2F) links by the inter-Fog physical wireless backbone. Container-based virtualization is used to make a virtual clone associated with physical things. The virtualization layer supports efficient use of limited resources and generates the virtual clone of physical things. The Fog node physical server serves cloned things and an overlay inter-clone virtual network was established which allows P2P communication among clones by depending on TCP/IP end to end transport connections.

\subsubsection{Fog computing Architecture based on nervous system} 
Sun and Zhang \cite{sun2017resource} presented an architecture that was constructed based on the human nervous system. In their proposed architecture, the cloud data centre is considered as the brain nerve centre, the Fog computing data centre is considered as the spinal nerve centre and smart devices are considered as peripheral nerve centres. These three nerve centres spread their connections extensively throughout all of the body of the system. Peripheral nerves scattered in the body and the brain will control all the activities of the spinal cord. The structure of the system is designed based on the neural structure of the human body where the brain is responsible in dealing with all tasks. All the smart devices connected to this system are referred to as the peripheral nerves which are geographically distributed. These devices include tablets, phones, sensors, or smart watches. The Fog computing centre would address certain simple and time-sensitive requests, for example, the spinal cord knee jerk reflex can share the resource pressure of the cloud data centre. The spinal cord is the connecting route between the brain and peripheral nerves; this is alike to the location of the fog data centre that joins the Internet of Things with high-level cloud data centres. \\ 

\begin{figure}[!b]
	\centering
	\includegraphics[width=3.8in]{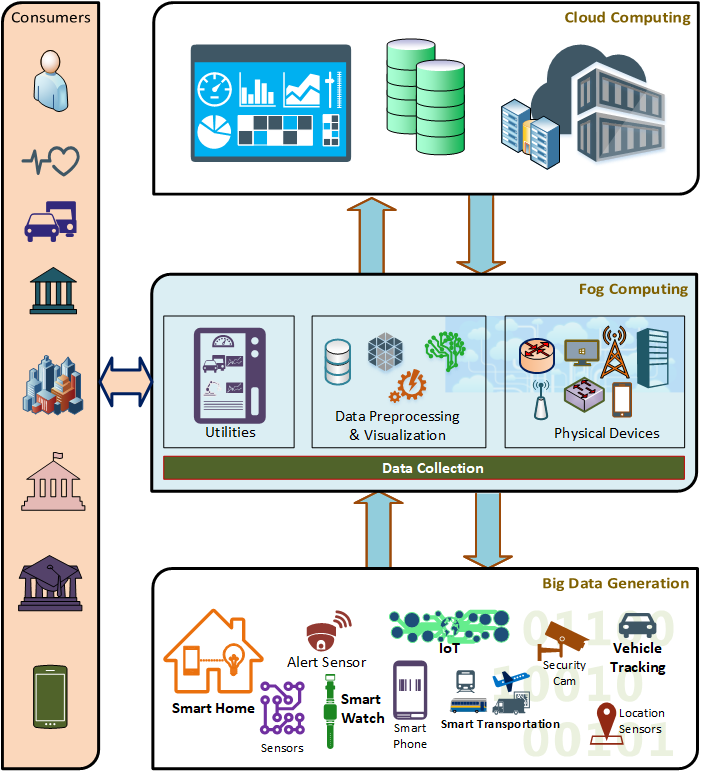}
	%where an .eps filename suffix will be assumed under latex, 
	%and a .pdf suffix will be assumed for pdflatex; or what has been declared
	%via \DeclareGraphicsExtensions.
	\caption{Layered architecture of Fog Computing.}
	\label{og_fog_hl_arc}
\end{figure}

\subsubsection{IFCIoT architecture} Integrated Fog cloud IoT (IFCIoT) architecture has been proposed by Munir et al. \cite{munir2017ifciot}. This architecture enables federated cloud services for IoT devices through an intermediary Fog infrastructure. The federated cloud is formed by multiple external and internal cloud servers which match application and business needs. Gateway devices, smart routers, edge servers and base stations are the Fog nodes and much of the processing takes place in these nodes. Fog nodes are autonomous; thus, each node can ensure uninterrupted service by their providers. The entire deployment of the fog computing environment can be local deployment in case of automation of single office buildings and can also be distributed in the regional level, including local levels in the case of large commercial companies located in multiple buildings in various places in the IFCIoT architecture. This architecture supports distributed deployment and information from various levels feed to a centralized system. The connectivity of all IoT devices is considered to be wireless connectivity via WLAN, WiMAX, and other cellular networks. Fog nodes maintain a connection with IoT devices within its wireless range. The entire Fog is connected to the federated cloud service through the core network. For collaborative processing, a Fog system could be connected to other Fogs wirelessly. \\

{\renewcommand{\arraystretch}{1.5}
\begin{sidewaystable}[htbp]
	\centering
	\caption{Features focused by various architectures.}
	\label{arcsum}
	\begin{tabular}{L{4.5cm}|L{0.6cm}C{0.6cm}C{0.6cm}C{0.6cm}C{0.6cm}C{0.6cm}C{0.6cm}C{0.6cm}C{0.6cm}C{0.6cm}C{0.6cm}C{0.6cm}C{0.6cm}C{0.6cm}} \toprule
		\textbf{Authors \& Year}&\textbf{{\rotatebox{90}{Physical Fog Devices}}}&\textbf{\rotatebox{90}{Virtualized Fog devices}}&\textbf{\rotatebox{90}{Fog Server}}&\textbf{\rotatebox{90}{Monitoring}}&\textbf{\rotatebox{90}{Energy Efficiency}}&\textbf{\rotatebox{90}{Data Preprocessing}} &\textbf{\rotatebox{90}{Temporary Storage}}&\textbf{\rotatebox{90}{Security \& Privacy}}&\textbf{\rotatebox{90}{Cloud Storage}}&\textbf{\rotatebox{90}{Scalability}}&\textbf{\rotatebox{90}{Autonomy}}&\textbf{\rotatebox{90}{Programmability}}&\textbf{\rotatebox{90}{Architecture Validation}}&\textbf{\rotatebox{90}{Reliability}}\\ \hline
		
		Aazam et al. (2015)\cite{aazam2015fog} &\ding{51}&\ding{51}&\ding{55}&\ding{51}&\ding{51}&\ding{51}&\ding{51}&\ding{51}&\ding{51}&\ding{55}&\ding{55}&\ding{55}&\ding{51} &\ding{55}\\ \hline
		
		Luan et al. (2015) \cite{luan2015fog} &\ding{51}&\ding{55}&\ding{51}&\ding{55}&\ding{55}&\ding{51}&\ding{51}&\ding{55}&\ding{51}&\ding{55}&\ding{55}&\ding{55}&\ding{55} &\ding{55}\\ \hline
		
		Giang et al. (2015) \cite{giang2015developing}
		&\ding{51}&\ding{55}&\ding{55}&\ding{55}&\ding{55}&\ding{51}&\ding{51}&\ding{55}&\ding{51}&\ding{51}&\ding{55}&\ding{51}&\ding{51} &\ding{55}\\ \hline
		
		Dastjerdi et al. (2016) \cite{dastjerdi2016fog}
		&\ding{51}&\ding{55}&\ding{55}&\ding{51}&\ding{55}&\ding{51}&\ding{51}&\ding{51}&\ding{51}&\ding{51}&\ding{55}&\ding{55}&\ding{51} &\ding{55}\\ \hline
		
		OpenFog (2016) \cite{openfog2016openfog}
		&\ding{51}&\ding{51}&\ding{55}&\ding{51}&\ding{55}&\ding{51}&\ding{51}&\ding{51}&\ding{51}&\ding{51}&\ding{51}&\ding{51}&\ding{55} &\ding{55}\\ \hline
		
		Arkian et al. (2017) \cite{arkian2017mist}
		&\ding{51}&\ding{51}&\ding{55}&\ding{55}&\ding{51}&\ding{51}&\ding{51}&\ding{51}&\ding{51}&\ding{55}&\ding{55}&\ding{51}&\ding{51} &\ding{55}\\ \hline
		
		Baccarelli et al. (2017) \cite{baccarelli2017fog}
		&\ding{51}&\ding{55}&\ding{55}&\ding{51}&\ding{51}&\ding{51}&\ding{51}&\ding{55}&\ding{51}&\ding{55}&\ding{55}&\ding{55}&\ding{51} &\ding{51}\\ \hline
		
		Sun et al. (2017) \cite{sun2017resource}
		&\ding{51}&\ding{55}&\ding{55}&\ding{55}&\ding{55}&\ding{51}&\ding{51}&\ding{55}&\ding{51}&\ding{55}&\ding{55}&\ding{55}&\ding{51} &\ding{55}\\ \hline
		
		Munir et al. (2017) \cite{munir2017ifciot}
		&\ding{51}&\ding{51}&\ding{51}&\ding{51}&\ding{51}&\ding{51}&\ding{51}&\ding{55}&\ding{51}&\ding{51}&\ding{55}&\ding{55}&\ding{55} &\ding{55}\\ 
		
		\botrule\end{tabular}
\end{sidewaystable}
}

Table \ref{arcsum} summarizes the supported features of Fog computing architectures proposed by various researchers in the past few years. According to this table, all proposed architectures represent the concept of data processing at the edge and usage of Fog devices for temporary storage whereas cloud infrastructure will be used for long-term storage. Most of these architectures do not focus on virtualized devices as the Fog device but rather represent the Fog device as just a physical device. However, Fog devices can be virtualized (e.g. Cisco UCS server) and non-virtualized (e.g. smartphones). As reported by \cite{openfog2016openfog}, scalability is the localized control, command and processing; orchestration and analytics; and avoidance of network taxes. Autonomy represents flexibility, cognition, agility and value of data. Programmability is designated by programmable hardware and software, virtualization, multi-tenant, and app fluidity. However, these features were overlooked by most of the reviewed architectures. Some proposed architectures were also not validated by any model or experimental environment \cite{luan2015fog,openfog2016openfog,munir2017ifciot}. Alternatively, many of them were validated by theoretical models, simulations or experimental evaluations. Table \ref{arcval} summarizes the proposed models, performance metrics and simulation tools used by them.

\begin{table*}[ht]
	\centering
	\caption{Models for Fog computing architecture evaluation.}
	\label{arcval}
	\begin{tabular}{L{2.5cm}|L{2.5cm}L{2.5cm}L{2.5cm}} \toprule
		\textbf{Authors \& Year}&\textbf{Proposed Model} &\textbf{Simulation Tools Used} &\textbf{Performance Metrics} \\ \hline
		
		Aazam et al. (2015)\cite{aazam2015fog} &Fog based IoT resource management & CloudSim 3.0.3 & resource prediction, resource allocation, and pricing \\ \hline
		
		Giang et al. (2015) \cite{giang2015developing} & Distributed Dataflow (DDF) programming & N/A & N/A \\ \hline
		
		Dastjerdi et al. (2016) \cite{dastjerdi2016fog} & Dag of query for incident detection  & CloudSim & Average tuple delay, core network usage \\ \hline
		
		Arkian et al. (2017) \cite{arkian2017mist} & Cost-efficient resource provisioning & Not Mentioned & Service Latency, Power Consumption, Cost \\ \hline
		
		Baccarelli et al. (2017) \cite{baccarelli2017fog} & V-FoE testbed & iFogSim (Extension of CloudSim) & Energy Consumption, RTT, Connection Failure \\ \hline
		
		Sun et al. (2017) \cite{sun2017resource} & Repeated game based resource-sharing model & Not Mentioned & SLA Violation, Completion Time \\ 
		
		\botrule\end{tabular}
\end{table*}

\subsection{High level Fog computing layered architecture}
As shown in the layered architecture in Figure \ref{og_fog_hl_arc}, end users are connected to the Fog layer and this layer is responsible for maintaining communication with the data generation and cloud computing layer. In Fog computing layered architecture, there are four different layers: consumer layer, cloud computing layer, fog computing layer and big data generation layer. This architecture is the high level architecture of Fog computing which separates the main actors of the Fog computing environment. The consumer layer represents groups of people and institutes who will use Fog computing services. Actual Fog computation processing will be done by the cloud and the Fog computing layer. Various groups of people or sectors can benefit from the Fog computing services. This includes individuals, health-care sectors, smart transportation systems, private organizations, government organizations, smart cities, academia and any other smart system that needs time-sensitive processing. The consumer groups mentioned above are connected with various sensors and data generation devices in various ways. Based on the necessity of the particular group, they also have strict time and deadline constraint application requests. Healthcare and transportation systems have the most sensitive applications in these cases. Although, the implementation of Fog computing will benefit everyone regardless of group and organization as the cloud has been doing currently. The functionality of three main layers of Fog architecture is presented below. \\

\subsubsection{Fog computing layer} 
The fog computing layer is the most important layer compared to the others since it maintains communication with all other layers. The consumer layer directly sends processing requests to this layer except communication with data generation and the cloud computing layer. This layer is responsible for data collection from the end devices. This layer will decide whether or not a processing or storage request needs to be sent to the cloud. Various utilities for various services will be contained by this layer. Small-scale data processing will be done in this layer with virtualization support. The maximum processing is considered as stream processing and needs to be done in an online manner without storing huge amounts of data. However, short-term storage and pre-processing will be done by this layer. All processing and storage related operations should be executed by any device that has processing power and storage capacity. These devices are generally known as Fog devices and Fog servers. It could be any type of device including routers, switches, access points, base stations, smartphone, servers, and hosts. 

Application processing from the consumers can be done without the Fog layer with help from the cloud. But, the problem is time-critical applications that cannot rely on the cloud because of the location of the cloud infrastructure. As an example: in a smart transportation system, we assume that if two alternate roads are available from a specific point and one road becomes congested resulting in it taking some time to update the status due to cloud processing, during this processing interval, some vehicles are directed to the congested road. As a consequence, the total travel time to reach the destination is increased by half an hour. It may not affect ordinary people that much, but what if it happens to an ambulance directed to that route? It might even cost a life in some cases. To avoid such kinds of inconvenience, we should not rely on the cloud to process sensor data. Other types of latency critical decisions are also possible in the smarter transportation system, such as dividing traffic into several routes during busy hours, natural disasters (rock falling on the road, extreme fog or rain etc.), as well as safety of pedestrians, animals, and cyclists. The only solution in these kinds of applications is to accomplish processing at the closest location near to the users. This is what the Fog layer actually does. \\

\subsubsection{Data generation layer} 
The data generation layer contains all devices and sensors from where Big Data can be generated. Big Data are generated in our everyday life. First of all, when we are at home, we are using various smart devices for communication, and many other sensors and actuators belong to lighting systems, music and entertainment systems and alarm sensors are generating data by following certain intervals. Secondly, our regular activities such as shopping, working out, office work and school work lead us to generate Big Data. Finally, our regular hits to various services also produces data that might have useful insight, for example, we need to know traffic updates, task lists, offers in the various shops, activities, and tasks of the school. In this way, data is generated at every moment. We also find many smart devices and sensors everywhere. Security cameras, roadside dash cams, speed cameras, temperature sensors, GPS sensors, alarm sensors and actuators are generating data every jiff as we keep moving. We are also using the smart device for tracking our exercise and other activities or even sleeping activities. Hence, all devices that can generate data belong to this layer and transfer generated data to the Fog computing layer. Any actionable request which is generally done by the actuator will be initiated from the Fog computing layer when necessary. Thus, this layer has two-way communication with the Fog computing layer. Data generated by the Big Data generation layer will be collected and analysed by the Fog computing layer. The Fog computing layer will decide whether storage and processing will be done by its own resources or if there is a need to use cloud resources. \\     
       
\subsubsection{Cloud computing layer} 
Traditional cloud infrastructure resides in this layer. The main functionality of this layer is to provide long-term storage and complex time insensitive processing. As this layer cannot communicate directly to the consumers and end devices, it completely depends on the Fog computing layer for communication. From the user’s perspective, users are no longer submitting their request to the cloud; they are submitting requests to the Fog computing layer. As a result, processing will be done more quickly when users are depending on Fog computing resources which are located in the close proximity. It is not possible to eliminate the cloud computing layer because it assumes that the devices located in the Fog computing layer have minimal computing and storage resources. Thus, if we need to store a high volume of data and process the same, we would have no other option but to depend on the cloud. 

\section{Limitation of the cloud to execute Big Data applications}

\subsection{Exploding generation of sensor data}
Due to the high pace of IoT technology adoption, Big Data generation is increasing excessively. It is expected that the total number of connected devices worldwide will be about 30 billion by 2020 and it will further increase to 80 billion by 2025 which is nearly triple within a five-year gap \cite{kanellos_2016}. It is also predicted that by 2025, about 152,000 new devices will be connected to the Internet every minute \cite{kanellos_2016}. According to an IDC report, by 2020, about 40\% of data will be processed at the edge \cite{idccomiot}. It is hard to send all generated data into the cloud due to communication costs. Hence, it is better to process the generated data near to the users without sending to the cloud. IoT has been crafted to create greater opportunities that can help reshape the modern world via increases in revenue and reductions in operational costs by revealing better insight where collections of huge data alone are not sufficient. To get the real benefit of IoT revolution, organizations need to develop a platform where it is possible to gather, manage and evaluate massive sensor-generated data in an efficient and cost-effective manner \cite{arkian2017mist}.

\subsection{Inefficient use of Network bandwidth}
The cloud is the best place to process Big Data because it has high computation and processing resources. But sending all generated data to the cloud for processing will significantly increase network traffic. Hence, it is necessary to reduce the volume of data at the edge as well as to mine data to find the pattern at the edge level. Some applications, video surveillance systems for example, generate a high volume of video data. There are two critical issues arising from such systems. Firstly, if we send all video data to the cloud, it may occupy the maximum available bandwidth before sending them to the cloud. Secondly, processing these data at the edge is also challenging because it needs a large amount of processing power. As we mentioned earlier, the cloud is the best for the processing of high volumes of data, but the problem is that we do not have unlimited available bandwidth. The complexity of such systems will be increased day by day while video data are important for crime control and the implementation of a smart monitoring system in a smart city. So, the cloud is not the convenient way to deal with the applications that need to process video files.    

\subsection{Latency awareness}
Applications related to augmented reality, online gaming, smart homes and smart traffic are more latency sensitive. Fog nodes are usually located one or two hop distances from the user. Hence, if the Fog concept is used for Big Data processing at the edge, it can easily support latency aware applications. In contrast, the cloud is located in a multiple hop distance which is quite far from the user where latency is higher than Fog. Latency also affects online businesses. As examples, a 100ms delay may cause a 1\% reduction in Amazon sales and a 500ms delay would cause a 20\% drop in Google traffic \cite{greenberg2009networking}. 

\subsection{Location awareness}
Most IoT applications are context aware, meaning that application processing depends on the location and other applications running nearby. Sending all these context-aware application requests to the cloud is not realistic and sometimes not affordable due to bandwidth and delay constraints. Many IoT applications also require minimum processing delays of less than a number of milliseconds \cite{zhang2017survey} such as health-care applications, vehicular networks, drone control applications and emergency response systems. Such applications also require real time data processing. These types of application services always depend on the surrounding environmental data rather than information available to other locations. 

Location awareness is also used in smart traffic applications to detect the pattern of the traffic such as roadwork, roadblocks, traffic congestion and accidents. These applications share information among connected vehicles to improve vehicle navigation and traffic management \cite{koldehofe2012moving}. Another example is the smart surveillance application system where a police officer of a local police station can see the video stream of suspicious people around his designated area and is able to track and initialize action to prevent damage to the public \cite{hong2011target}. In the above scenarios, processing and running applications on the cloud may not efficient enough due to the high response times of the cloud.

\section{Challenges faced when executing Big Data applications on Fog}
Data pre-processing and post-processing is completely dependent on the cloud while we depend on the cloud for Big Data processing. On the other hand, in Fog, Big Data streams are pre-processed locally by using a number of spatially distributed stationary or mobile devices. Then, the processed data are sent to the cloud data centre for further post-processing while necessary. Pre-processing, transformation, and post-processing can be referred to as a three-phase life cycle of data processing in Fog. Fog devices can be mobile and they also have resource-limited properties, therefore, there are many obstacles that exist when executing Big Data applications on Fog. This section points out the key issues of Big Data application execution on Fog.      

\subsection{Resource limited Fog device}
Fog devices used for computation have limited resources for processing. Compared to the cloud, the resources of Fog devices are very limited. However, Big Data can be executed on these devices by making Fog clusters as Google uses commodity machines as the server and also introduced the map-reduce concept to deal with huge data processing using commodity machines. Any device that has computation capabilities can act as a Fog device, from user devices to all kinds of network management devices are considered as Fog devices. Generally, these devices have their own operating systems and applications that occupy most of the system resources. Running Fog applications on these devices is always the second priority. Thus, efficient and intelligent job allocation policy is required in order to protect a high number of job failures.          

\subsection{Power Limitation}
First of all, in the Fog paradigm, any device that has processing, storage and network capabilities can act as a Fog device. Hence, battery-powered mobile devices such as laptops, smartphones and tablets can act as a Fog device \cite{naha2018fog}. The energy of these devices is limited and due to this, the device may turn off from low battery levels during computation. As a consequence, tasks that were running before power off need to be rescheduled to another device. Thus, the scheduler might aware of power availability before task allocation to a mobile device. Secondly, many types of sensor devices are powered by batteries and these batteries should be recharged frequently. It is a challenging issue in some cases as failure due to the battery may cause serious harm. For example, body sensor networks which are monitoring patients. In such scenarios, we should use energy efficient methods by incorporating primary data analytics which will reduce the volume of data that needs to be stored and transmitted \cite{dubey2015fog}. 

\subsection{Selection of Master node}
In the Fog environment, many of the devices might be mobile and these devices are always moving which may cause a problem to make the cluster of Fog devices. Big Data processing using map-reduce where it is necessary to select a master node is mainly responsible for tracking jobs. If the master node moved to another cluster, then it will be necessary to execute the whole job again. To address this issue of master node selection in the Fog cluster is challenging. However, selecting a master node among stationary nodes is a good solution. 

\subsection{Connectivity}
Most of the Fog and IoT devices are connected using wireless connectivity which has problems with interference and fading. These problems cause fluctuations of the access bandwidth. Beside this, in the Fog environment thousands of devices and sensors coexist and communicate with each other repeatedly. Also, the number of connected devices will be increasing over time. To address this issue, it is crucial to develop an efficient model that can estimate the number of connected devices within a Fog device and can predict resource incorporation before the failure of the system.    

\section{Recent advances on Big Data application execution on Fog}
Fog computing is a comparatively recent research trend and much research work has been done in this area. However, only a few research works have addressed specifically Big Data application execution on the Fog paradigm. Most of the work has been focused on the healthcare system and also some work has been done in various areas including smart cities and virtual learning systems.

Dubey et al. \cite{dubey2015fog} proposed a service-oriented architecture for telehealth Big Data in Fog computing by adding low powered embedded computers in the Fog layer which performs data analytics and data mining on raw data. They considered two use cases, speech motor disorder and cardiovascular detection. Raw data for both use cases were processed in the Fog layer. After the processing of raw data from cardiovascular detection sensors or speech motor disorder sensors, all detected patterns are stored and the unique, distinctive pattern is sent to the cloud for further processing. In the first case study, speech data was sent to the Fog device by the smartwatch. Then, the Fog device does three steps of processing, feature extraction, pattern mining and compression. In this way, it converts speech into average fundamental frequency and loudness features. The compressed speech data is then sent to the cloud. The cloud processes the received average fundamental frequency and loudness to convert it into original speech time-series. For this case study, Dynamic Time Warping (DTW) was used for speech data mining and Clinical Speech Processing Chain (CLIP) algorithm was used for computing relevant clinical parameters such as fundamental frequency and loudness. In the other case study on ECG monitoring, DTW was used for pattern recognition, and after recognizing the pattern, the processed data was sent to the cloud for further processing.  

Ahmad et al. \cite{Ahmad2016} proposed a framework health monitoring in the Fog paradigm where the Fog is an intermediate layer between cloud and users. The proposed design feature reduced communication costs compared to other available systems. The work also proposed utilizing the Cloud Access Security Broker (CASB) to deal with security and privacy issues. The sensory data will be generated by a smartphone in 3 second intervals and will be sent to a local machine known as the Health Fog in every minute interval in a batch. Besides activity data, other sensory data from smart homes and hospital activities will be stored in the Health Fog. Then, the intermediate processing will be done on Fog and shared with nutritionists or doctors as per user preference. The final processed information on calorie burning and activity detection will be deposited in the cloud and shared according to user preference.

Tang et al. \cite{tang2017incorporating,tang2015hierarchical} studied smart pipeline monitoring in smart cities to find threatening events by using a sequential learning algorithm which is based on using fibre optic sensors to analyse Big Data in Fog computing. The authors have presented 4-Layer architecture for monitoring the pipeline smartly and it can be used for other infrastructures like smart buildings and smart traffic. The lowest layer of their architecture is the fibre optic sensor network where the cross-correlation method and time-domain filter is used to detect changes in the physical environment such as stress, temperature and strain. The next upper layer consists of small computing nodes working in parallel. Each node is responsible to perform two tasks, the first is to detect potential threat patterns from sensor data using machine learning algorithms and the second is the feature extraction to do the further processing by the upper layer known as intermediate computing nodes. To reduce communication load, raw data from sensors will not be received by the intermediate computing nodes. This layer coordinates data analysis from various locations to identify hazardous events in a specific region. The most upper layer is the cloud which is basically built by using Hadoop clusters to determine and predict long-term natural disasters. The hidden Markov model is used as a sequential learning method and has been verified successfully in a real environment.

Zhang et al. \cite{Zhang2017} presented a hierarchical model for resource management in inter and intra Fog environment which considered packet loss and energy efficiency of the intelligent vehicle and Fog server. The model is comprised of two layers: Fog layer and edge layer. The Fog layer is the association of local Fog servers, clouds, and coordination servers. On the other hand, Vehicular Ad-hoc Networks (VANETs), cellular networks and IoT applications are the key elements of the edge layer. The Intra Fog resource management manages internal tasks assigned to the virtual machine within the Fog server and the task size is ordered by an Adaptive Load Dispatcher (ALD) where the virtual machine processing rate is adjustable. In inter Fog resource management operations, all local Fog servers update their working states to a server, which coordinates all Fog servers and assigns overflow workload to idle Fog servers nearby. This inter-fog resource management operation controls the data flow with the help of access control routers without disturbing intra-fog operations.

A distributed resource sharing scheme is proposed by Yin et al. \cite{Yin2017} where the Software Defined Network (SDN) controller dynamically adjusts the quantity of application data that will be pre-processed by the Fog nodes for Big Data streaming applications. The problem of application data quantity assignment is formulated as a social welfare maximization problem in their work. Also, the loss of information value occurring in the pre-processing process was contemplated. A Hybrid Alternating Direction Method of Multipliers (H-ADMM) algorithm was employed to solve computation burdens in a fully distributed environment composed of Fog nodes, SDN controllers and cloud. Moreover, an efficient message exchange pattern was integrated to reduce communication costs on the SDN controller when it’s dealing with a large number of Fog nodes.

Pecori \cite{Pecori2018} proposed a virtual learning architecture where Big Data streams were processed in a Fog computing environment using Apache Storm and Scalable Advanced Massive Online Analysis (SAMOA) as distributed machine learning framework. The Storm real-time computation system was chosen for implementation due to its inbuilt master-slave architecture that can be mirrored easily between Fog and cloud. The reason behind choosing Storm instead of the spark is that the Storm is event-oriented rather than batching data updates at a routine interval. The SAMOA was selected for the deployment of its finer integration with Storm. In the proposed architecture, cloud infrastructure was used for high volume storage, historical backup operation and mining jobs with high latency. The cloud also provides long-term forecasts and scores of selected features by communicating with intermediate and macro users such as educational institution managers and the policy makers through outer APIs. The Fog layer is composed of lightweight storage facilities with distributed NoSQL support along with multiple Storm slaves. These slaves can be network gateways, sensors or any other smart devices and they can perform short-term predictions by using light mining techniques. These predictions provide useful suggestions to all consumers, students, tutors and teachers in an e-learning environment. Using the above proposed technique, the Fog based stream processing virtual learning framework would be beneficial for students and instructors as well as distance learning institutions.

A summary of the above Fog based Big Data application frameworks and architectures are presented in Table \ref{bdfog} and Table \ref{bdfog1} with their limitations.

{\renewcommand{\arraystretch}{1.5}
\begin{sidewaystable}[htbp]
	\centering
	\caption{Research works that focused on both Fog and Big Data.}
	\label{bdfog}
	\begin{tabular}
    {L{2.5cm}|C{2cm}C{1.5cm}C{2cm}C{1.5cm}C{2cm}C{2cm}C{2cm}} \hline \toprule
		\textbf{Authors \& Year}&\textbf{Address Big Data Applications}&\textbf{Fog Computing Paradigm}&\textbf{Application Type}&\textbf{Layers in Architecture}&\textbf{Distributed Processing Framework for Cloud}&\textbf{Distributed Processing Framework for Fog} &\textbf{Devices used as Fog}\\ \hline
		
		Tang et al. (2015, 2017)\cite{tang2015hierarchical,tang2017incorporating} &Yes&Yes&Smart Cities&4-Layers&Hadoop&Not mentioned&Not mentioned\\ \hline
        
        Dubey et al. (2015)\cite{dubey2015fog} &Yes&Yes&Telehealth&3-Layers&Not mentioned&Not mentioned&Intel Edison\\ \hline

        Ahmad et al. (2016)\cite{Ahmad2016} &Yes&Yes&Healthcare&3-Layers&Not mentioned&Not mentioned&Windows based machine\\ \hline 
        
        Zhang et al. (2017)\cite{Zhang2017} &Yes&Yes&Smart cities&2-Layers&NA&NA&NA\\ \hline
        
        Yin et al. (2017)\cite{Yin2017} &Yes&Yes&Data assignment to each Fog node&3-Layers&NA&NA&NA\\ \hline
        
        Pecori (2018)\cite{Pecori2018} &Yes&Yes&Visual learning&3-Layers&Storm&Storm&Smartphone\\ \hline

		\botrule\end{tabular}
\end{sidewaystable}
}

{\renewcommand{\arraystretch}{1.3}
\begin{sidewaystable}[htbp]
	\centering
	\caption{Algorithms, architecture, framework and limitation of various research works on Fog computing for Big Data.}
	\label{bdfog1}
	\begin{tabular}
    {L{3.5cm}|C{4cm}C{4cm}C{4cm}} \hline \toprule
		\textbf{Authors \& Year}&\textbf{Algorithms Used}&\textbf{Proposed Architecture or Framework}&\textbf{Limitation}\\ \hline
		
		Tang et al. (2015, 2017)\cite{tang2015hierarchical,tang2017incorporating} &Sequential learning algorithms&Big Data analysis architecture in Fog computing &	Fixed speed of computing nodes and did not consider memory access time\\ \hline
        
        Dubey et al. (2015)\cite{dubey2015fog} &Dynamic Time Warping (DTW)
and Clinical Speech Processing Chain (CLIP)	&Service oriented architecture for Fog computing	& Complex speech analysis to recognize more accurate speech disorder \\ \hline

        Ahmad et al. (2016)\cite{Ahmad2016} &Homomorphic encryption	& Framework for health and wellness applications	& Distributed Fog environment is not present\\ \hline 
        
        Zhang et al. (2017)\cite{Zhang2017} &NA&Hierarchical resource management model for intra and inter Fog &	Adoption of the model in real environment\\ \hline
        
        Yin et al. (2017)\cite{Yin2017} &Hybrid Alternating Direction Method of Multipliers (H-ADMM) algorithm	& Distributed resource sharing scheme with SDN controller	& Does not consider incentive mechanism\\ \hline
        
        Pecori (2018)\cite{Pecori2018} &Distributed Hash Tables (DHT)
and Machine learning algorithms &	e-learning architecture with the integration of Cloud & Fog and Big Data	Overlooked the security and privacy concern\\ \hline

		\botrule\end{tabular}
\end{sidewaystable}
}

%&\textbf{

\section{Fog computing products}
\subsection{Cisco IOx}
The Cisco IOx application platform enables Fog processing near to the end devices by integrating IoT sensors and the cloud. Cisco IOx is the combination of four components: IOS and Linux OS, Fog Director, the SDK and development tools, and Fog applications. Cisco IOS is the leading NOS which ensures secure network connectivity and Linux is an open source customizable platform. However, this platform is designed to execute applications on the Cisco IoT network infrastructure. Fog Director helps to administrate applications over the network running on the Cisco IOx environment. The SDK and development tools provide a collection of tools, command line utility and web-based applications for the developers. Fog applications are readily executable applications that will run on IOx enabled infrastructure \cite{ciscoiox2016}. Cisco 800 series routers support IOx and it can be used as a Fog device because it supports two OS on two cores. One core dedicated to IOS and the other core supports Linux based OS which helps to do the processing in the fog environment \cite{cisco_2016IOx}.

\subsection{LocalGrid's Fog computing platform}
This is an embedded software that can be installed on network edge devices. It enables various communication protocols into an open standard by the LocalGrid Fog computing platform. The embedded software can be installed on all sensors, switches, routers, machines and other edge devices which allow secure access for Fog processing and also leverage the communication gaps between new and legacy devices. It also forms a P2P communication between multiple edge devices and facilitates real-time coordination and control without a centralized server. The LocalGrid also has a communication infrastructure with the cloud by incorporating intelligence network communication which reduces latency, improves security and reduces usage of bandwidth. On this platform, data processing and decision making on edge devices can be done on a distributed way  \cite{localgrid}.

\subsection{Fog Device and Gateways}  Besides proprietary products, there are many other alternatives that can be used for Fog devices and gateways commonly known as a computer on modules. These devices include Intel Edison, Raspberry Pi, Arduino, Asus Tinker Board, Odroid-C2, Banana Pi M2 Ultra, and Odroid-XU4. Among them, the Raspberry Pi is most popular. However, we may find many other vendors whose are producing more powerful devices than the Raspberry Pi ata competitive price. Most of these devices support Android OS and can use other types of OS, including Raspbian, OSMC, OpenELEC, Windows IoT Core, RISC OS and Ubuntu MATE. Many research studies have been done using these computers on modules \cite{dubey2015fog,barik2017fog2fog,barik2018mist, hajji2016understanding, he2017multi,giordano2016smart} in the past couple of years to implement the Fog environment for Big Data processing. 

\section{Research Issues}
\begin{figure}[!t]
	\centering
	\includegraphics[width=4in]{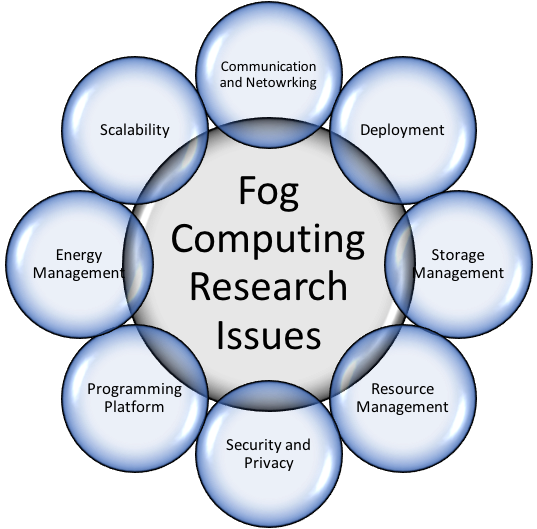}
	%where an .eps filename suffix will be assumed under latex, 
	%and a .pdf suffix will be assumed for pdflatex; or what has been declared
	%via \DeclareGraphicsExtensions.
	\caption{Research issues of Fog computing for Big Data application execution.}
	\label{bdrisfog}
\end{figure}

There are several research issues currently available that need to be addressed for Big Data application execution on Fog. Figure \ref{bdrisfog} represents several research issues in this area.

To run Big Data IoT applications in Fog, stream processing is done in the Fog environment and Big Data processing will be done in the cloud. However, it is possible to implement Fog cluster near to users using Fog devices and have potential to execute Big Data related processing to some extent. To do so, one of the key issues is cross-layer communication between the Fog to the cloud and the end user to the Fog. Besides cross-layer communication, inter Fog communication is also equally important. In inter Fog communication, data transmission is contested by heterogeneous service policies, network topology and device connections \cite{luan2015fog}. Moreover, integration of emerging technologies such as 5G technologies, Software Defined Networking (SDN) and Network Function Virtualization (NFV) is also necessary to incorporate these technologies with Fog. 

With regards to the deployment, the main challenges are application placements and scaling. First of all, network operators need to customize the applications based on local demand. Secondly, due resource-limited properties of Fog devices, it is difficult to scale resources as per user demand. Thirdly, since the location is more important in Fog environment, placing Fog servers in the right place is also challenging \cite{luan2015fog}. 

To satisfy low-latency requirements, pre-cache technology needs to be employed and Fog nodes should store cache to the Geo-distributed nodes by passively predicting users’ demands. In this way, delays when downloading content will be reduced significantly and efficient use of storage resources fulfils Fog application requirements. The storage resource of Fog device is limited so storage expansion technologies are very effective in order to improve overall service quality \cite{hu2017survey}.

Resource management in Fog has the utmost priority to maintain the lifespan and performance. Hence, resource scheduling techniques including placement, consolidation, and migration need to be investigated extensively \cite{hu2017survey}. Application placement to these resources is another research challenge for time-critical computing applications such as emergency response, healthcare, vehicular network, augmented reality, online gaming, brain-machine interface and any other smart environment related applications. Besides the above challenges, energy management, programming platforms, security, privacy, and scalability are also important research issues. Also user's privacy \cite{aghasian2018user} is most important that need to be explore in Fog perspective. 

\section{Conclusion}
Fog computing is an emerging technology which has flourished in solving Big Data IoT application execution problems by processing continuously generated data at the edge. This computing paradigm is a high-potential computing model that is growing rapidly but it is not mature enough as many issues still need to be investigated extensively. This paper reviewed and presented several existing architecture of Fog computing in order to identify the research issues related to Big Data application execution using Fog paradigm. We presented a high-level Fog computing architecture and discussed many other architectures of Fog computing and highlighted the features of numerous proposed architectures. We also discussed key limitations of the cloud to execute Big Data applications, especially in the IoT environment. Following the limitations of cloud, some challenges to execute Big Data application on Fog were presented. Also, some recent research works that specifically addressed Big Data application executions on Fog were investigated. Consequently, the characteristics of some currently available commercial Fog related platforms and devices were discussed. Finally, several open research issues were presented. Hopefully, these will pave future research directions among industry experts and academia.

\bibliographystyle{vancouver-modified}
\bibliography{sample-vancouver}

\end{document}